\documentclass[a4paper,11pt]{article}
\usepackage{jcappub} 
\usepackage{lineno}

\newcommand{\LMaxTeV}{L_{\rm TeV,\, Max}}

\newcommand{\LMax}{L_{\rm Max}}
\newcommand{\FMax}{F_{\rm Max}}
\newcommand{\LMin}{L_{\rm Min}}

\arxivnumber{2306.16305} 
\title{\boldmath Setting an upper limit for the total TeV neutrino flux from the disk of our Galaxy}

\author[a,b,1]{V. Vecchiotti\note{Corresponding author.}}
\author[c,b]{and F.L. Villante}
\author[b]{and G. Pagliaroli}
\affiliation[a]{Department of Physics, NTNU,\\
NO-7491 Trondheim, Norway}
\affiliation[b]{INFN, Laboratori Nazionali del Gran Sasso,\\
67100 Assergi (AQ),  Italy}
\affiliation[c]{Department of Physical and Chemical Sciences, University of L'Aquila, \\
67100 L'Aquila, Italy}

\emailAdd{vittoria.vecchiotti@ntnu.no}

\abstract{We set an upper limit for the total TeV neutrino flux expected from the disk of our Galaxy in the region $|l|<30^{\circ}$ and $|b|<2^{\circ}$ probed by the ANTARES experiment.
We include both the diffuse emission, due to the interaction of cosmic rays with the interstellar medium, and the possible contribution produced by gamma-ray Galactic sources.
The neutrino diffuse emission is calculated under different assumptions for the cosmic ray spatial and energy distribution in our Galaxy.
The source contribution is instead constrained by analysis of the gamma-ray TeV sources included in the H.G.P.S. catalog.
In particular, we assume that the total gamma-ray flux produced by all the sources, resolved and unresolved by H.E.S.S., is produced via hadronic interaction and, hence, is coupled with neutrino emission.
We compare our total neutrino flux with the recent ANTARES measurement of the neutrino from the Galactic Ridge.
We show that the ANTARES best-fit flux requires the existence of a large source component, close to or even larger than the most optimistic predictions obtained with our approach.}

\begin{document}
\maketitle
\flushbottom

\section{Introduction} \label{sec:outline}

Neutrino telescopes have finally reached the required sensitivity to probe TeV neutrino production in our Galaxy.
During the past few years, several upper limits were obtained for Galactic ridge TeV neutrino emission. 
By considering the region $|l|<40^\circ$, $|b|<3^\circ$, ANTARES reported in, e.g., \cite{ANTARES:2016mwq} an upper bound for the one-flavor neutrino flux at the level of $6.0 \times 10^{-5} (E_\nu/1\,{\rm GeV})^{-\Gamma_\nu}\, {\rm GeV}^{-1}\, {\rm cm}^{-2}\,{\rm s}^{-1}\,{\rm sr}^{-1}$ for the assumed neutrino spectral index $\Gamma_\nu = 2.5$.
Other restrictive limits were also obtained, by using different techniques and/or classes of data by ANTARES \citep{ANTARES:2017nlh}, IceCube \citep{IceCube:2017trr} and by a joint analysis performed by the two experiments \citep{ANTARES:2018nyb}.  
More recently, a possible breakthrough was achieved by ANTARES collaboration that has reported the first possible hint of neutrino emission from the Galactic ridge  with $2.2\,\sigma$ significance in the angular region $|l|<30^{\circ}$ and $|b|<2^{\circ}$ and in the $1-100$ TeV energy band \citep{ANTARES:2022izu}\footnote{After the conclusion of this work, IceCube reported the observation of high-energy neutrinos from the Galactic plane \citep{IceCubeScience} at 4.5$\sigma$ level significance, see note added.}.
In this experimental context, it is particularly relevant to discuss the expected contributions to TeV Galactic neutrino signal on a quantitative basis and to interpret the ANTARES result in light of these expectations.


A guaranteed contribution to the Galactic neutrino signal is provided by diffuse cosmic rays (CRs) interacting with the interstellar medium (ISM).
This component, called diffuse neutrino emission,  was predicted and discussed by several recent works and can be calculated in the TeV energy domain, with
consistent results, both by using complex CR propagation codes, such as GALPROP \citep{Galprop} DRAGON \citep{Evoli:2008dv,Evoli:2016xgn}, etc., or by using a "bottom-up'' approach where CR space and energy distribution in the
Galaxy is more directly linked to the local CR properties, see, e.g., \cite{Pagliaroli:2016,Cataldo:2019qnz,Lipari:2018gzn,Schwefer:2022zly}.
It was shown e.g. by \cite{Pagliaroli:2016} that diffuse neutrinos are expected to provide a non-negligible and potentially observable contribution to the IceCube data.
It is useful to recall that the same mechanism that produces the diffuse neutrino emission, also produces a diffuse gamma-ray background.
In this respect, the recent determination of diffuse gamma-ray flux at sub-PeV energies by Tibet AS$\gamma$ \citep{TibetASgamma:2021tpz} and LHAASO-KM2A \citep{LHAASO:2023gne} indicates that diffuse neutrino component should extend to very high energies.  
The interpretation of experimental results is, however, not straightforward because of the possible contamination of the observed gamma-ray signal by unresolved sources that can be quite large at these energies, as it is, e.g., discussed in \cite{Vecchiotti:Tibet}.


In addition to diffuse emission, neutrinos can be also produced by CR collisions within or close to their acceleration sites.
We refer to this component as the source contribution.
It is naturally expected that sources can produce a cumulative flux that is comparable to or larger than the diffuse neutrino emission at TeV energies.
In order to constrain this contribution, we can rely on TeV gamma-ray source catalogs like, e.g., the recently released H.E.S.S. Galactic Plane Survey (HGPS) \citep{H.E.S.S.:2018zkf}, which encompasses a large part of the Galactic disk.
However, gamma rays can be produced both by hadronic (via $\pi_0$ decay) and by leptonic interactions.
While the first mechanism predicts a comparable production of neutrinos (via decay of charged pions), the second one only accounts for photon emission.
In our Galaxy, there are different kinds of sources producing gamma rays above TeV energy.
The most abundant source class seems to be that of Pulsar Wind Nebulae (PWNe), see, e.g., \cite{Sudoh:2019lav,Abdalla:2017vci}.
The observed emission is believed to be produced by accelerated electrons and positrons through Inverse Compton (IC) mechanism. 
However, some studies suggest that PWNe are also able to accelerate a small fraction of protons \citep{Amato:2003kw}.
In this scenario, we cannot exclude that they also produce a subdominant neutrino flux.
In addition to PWNe, recent observations of Geminga and PSR B0656+14 by Milagro \cite{Abdo:2009ku} and HAWC \cite{Abeysekara:2017old} provided evidence for a new class of objects powered by pulsar activity, the so-called TeV halos, that could potentially explain a large fraction of bright TeV sources observed in the sky \cite{Sudoh:2019lav}.
Other important classes of Galactic sources are Supernova Remnants (SNRs) and star clusters.
Contrary to PWNe, these objects are believed to produce gamma rays via hadronic interactions.
Star clusters have been invoked as sources of high-energy CR in order to explain the CR observed composition \citep{Gupta:2019wmc}.
Recently, star clusters such as Westerlund $1$ \citep{Abramowski2012}, Westerlund $2$ \citep{Yang:2017arc}, Cygnus cocoon \citep{Ackermann2011Sci,Aharonian:2018oau} have been detected in gamma rays by different experiments, confirming that they are able to accelerate CR up to very high energy.
The model by \cite{Morlino:2021xpo} shows that the accumulated thin shell at the boundary of these objects could be the perfect environment for gamma-ray and neutrino production.

Since it is still largely debated which class of objects dominates the TeV sky and by which mechanism gamma-rays are predominately produced, in this paper we take an agnostic attitude and derive a conservative upper limit on the source neutrino component by assuming that TeV gamma-ray sources in the Galaxy are all powered by hadronic mechanisms.
We then take advantage of the fact that the Galactic source population can be efficiently constrained by fitting the HGPS gamma-ray source catalog as it is described by \cite{Cataldo:2020qla}.
This allows us to evaluate the maximal cumulative neutrino flux produced by all sources in our Galaxy.
We remark that our estimate, being based on a population study, not only accounts for observed sources but also includes by construction the contribution of sources that are not sufficiently luminous or sufficiently close to us to be resolved by gamma-ray experiments. 
The obtained upper limit is consequently very conservative and can be only evaded by assuming that the sources responsible for the neutrino production are opaque in gamma rays.

The plan of the paper is as follows. In Sect.~\ref{Diffuse neutrino emission}, we introduce our models to estimate the Galactic diffuse neutrino emission. In Sect.~\ref{Neutrinos from sources}, we calculate the maximal source contribution. In Sect.~\ref{Results}, we show our results and compare them with the ANTARES signal. In Sect.~\ref{Conclusions}, we draw our conclusions.

\section{Diffuse neutrino emission}
\label{Diffuse neutrino emission}

Accelerated protons and nuclei can escape the acceleration site, propagate in the Galactic magnetic field and interact with the ISM.
As a result, high-energy neutrinos and gamma rays are produced by hadronic interactions.
We calculate the neutrino diffuse flux by following the prescription of \cite{Pagliaroli:2016, Cataldo:2019qnz}. The differential one-flavor neutrino flux is obtained as:
\begin{eqnarray} \label{Neutrino diffuse flux}
\nonumber
\varphi_{\nu,{\rm diff}}(E_{\nu},\hat{n}_{\nu}) &=& 
\frac{1}{3} \sum_{l=e,\mu,\tau}  \int_{E_{\nu}}^{\infty} dE\; \frac{d \sigma_{l}(E,E_{\nu})}{dE_{\nu}} \\  
&&
\hspace{-1.5cm}
\int_{0}^{\infty} dl\; \varphi_{CR}(E ,r_{\odot} + l \hat{n}_{\nu})\, n_{\rm H}(r_{\odot} + l\hat{n}_{\nu})   
\end{eqnarray}
where $E_{\nu}$ and $\hat{n}_{\nu}$ indicate respectively the neutrino energy and arrival direction, while $\frac{d \sigma_{l}(E, E_{\nu})}{dE_{\nu}}$ represents the differential cross section for the production of neutrino and antineutrino with flavor $l$ by a nucleon of energy $E$ in a nucleon-nucleon collision. 
In the above relation, the neutrino flux at Earth is assumed to be equally distributed among the different flavors, as it is approximately expected due to neutrino mixing (see, e.g., \cite{Palladino:2015}). 

To compute the diffuse neutrino and gamma-ray fluxes, three fundamental ingredients are needed: the nucleon-nucleon cross-section, the number density of target nucleons $n_{\rm H}({\bf r})$ contained in the gas which is distributed along the Galactic disk, and the differential CR flux $\varphi_{CR}(E ,{\bf r})$ as a function of the energy and position in the Galaxy.
The cross-section is taken from \cite{Kelner:2006tc}.
The gas distribution, instead, is taken from the GALPROP code \citep{Galprop} and includes the contributions from atomic $\rm{H}$ and molecular $\rm{H}_{2}$ hydrogen.
The heavy element contribution is taken into account by assuming that the total mass of the interstellar gas is a factor 1.42 larger than the mass of hydrogen, as it is expected if the solar system composition is considered representative for the entire Galactic disk \citep{Ferriere:2001rg}.
The CR flux is parameterized as the product of three terms:
\begin{equation}
\varphi_{\rm CR}(E,{\bf r}) = \varphi_{\rm CR,\odot}(E)\,g({\bf r})\,h({E,\bf r})
\label{Eq:CR_flux}
\end{equation}
where $\varphi_{\rm CR,\odot}(E)$ represents the local nucleon flux, $g(r)$ describes the spatial distribution of CR and is an adimensional function (normalized to one at the Sun position ${\bf r}_\odot=8.5$ kpc).
The function $h({E,\bf r})$ introduces a position-dependent variation of the CR spectral index in order to reproduce the possible spectral hardening of large scale gamma-ray emission from the inner Galaxy that was recently inferred from analysis of the Fermi-LAT data \citep{Acero:2016qlg, Yang:2016jda, Pothast:2018bvh}.

The local CR nucleon flux $\varphi_{\rm CR,\odot}(E)$ is described according to the data-driven parameterization provided in \cite{Dembinski:2017} and it includes the contribution from all nuclear species.
In this work, we do not take into account the tension in the proton spectrum measured above $1$ PeV by KASCADE \citep{Apel:2013uni} and IceTop \citep{IceCube:2019hmk}. We checked, however, that the difference between the two experimental determinations may be responsible for small variations in our calculations and do not alter the conclusion of our work. 

The spatial distribution of CRs in our Galaxy $g({\bf r})$ is parameterized as the solution of a 3D isotropic diffusion equation with constant diffusion coefficient $D$ and stationary CR injection $f_{\rm S}({\bf r})$:
\begin{equation}
     \frac{\partial g({\bf r},t)}{\partial t} = \nabla^2 D\; g({\bf r},t)+f_{\rm S}({\bf r}).
     \label{Eq:g_funct}
\end{equation}
where $f_{\rm S}({\bf r})$ is assumed to follow the SNR number density parameterization given by \cite{Green:2015isa}. 
Hence, $g({\bf r})$ is defined as:
\begin{equation}
     g({\bf r}) = \frac{1}{\mathcal{N}}\;\int d^3 x\; 
     f_{\rm S}({\bf r}-{\bf x})\;
     \frac{{\mathcal F}(|{\bf x}|/R)}{2\pi |{\bf x}|}
     \label{Eq:g_funct1}
\end{equation} 
where $\mathcal{N}$ 
is a normalization constant:
\begin{equation}
     \mathcal{N} = \int d^3 x\; 
     f_{\rm S}({\bf r}_{\odot}-{\bf x})\;
     \frac{{\mathcal F}(|{\bf x}|/R)}{2\pi|{\bf x}|}
\end{equation} 
while the function ${\mathcal F}(\delta)$ is defined as:
\begin{equation}
  {\mathcal F}(\delta) \equiv \int_{\delta}^{\infty} d\gamma\; \frac{1}{\sqrt{2\pi}}\,\exp{\left(-{\gamma}^2/2\right)}
\end{equation} 
The parameter $R$ represents the diffusion length $R=\sqrt{2\,D\,t_{\rm G}}$ where $t_{\rm G}$ is the integration time. 
We assume two extreme values for this parameter, $R=1$ kpc and $R=\infty$, that allow us to reproduce the behavior of the CR density at $E\sim 20\,{\rm GeV}$ obtained by analysis of Fermi-LAT data, see \cite{Cataldo:2019qnz} for details. 
In the first case, CRs are confined close to their sources, and the spatial distribution resembles the one of SNRs \citep{Green:2015isa}, while in the second case,  CRs propagate in the whole Galaxy and the obtained spatial distribution is very close to that predicted by the GALPROP code \citep{Strong:2004td}. 

The function $h({E,\bf r})$ introduces the possibility that the CR spectral index is position-dependent. 
It is defined as:
\begin{equation}
h(E,{\bf r})=\left(\frac{E}{\overline{E}}\right)^{\Delta({\bf r})}
\label{Eq:h_funct}
\end{equation}
where $\overline{E}=20\,{\rm GeV}$  is the pivot energy and $\Delta({\bf r}_\odot)=0$. 
The function $\Delta({\bf r})$ in Galactic cylindrical coordinates is modeled as:
\begin{equation}
\Delta(r,z)=\Delta_0\left(1 - \frac{r}{r_{\odot}} \right)
\label{Eq:Delta} 
\end{equation} 
for $r\le 10$~kpc, while it is assumed to be constant for larger distances. The factor $\Delta_0=0.3$ represents the difference between CR spectral index at the Galactic center and its value at the Sun position.
This choice allows us to reproduce the spectral hardening of large-scale gamma-ray emission from the inner Galaxy that was recently inferred from Fermi-LAT data by \cite{Acero:2016qlg,Yang:2016jda,Pothast:2018bvh}
and is essentially equivalent to what was done by \cite{Lipari:2018gzn} in their "space-dependent" model. 
Moreover, it allows us to reproduce the predictions obtained by \cite{Gaggero:2014xla, Gaggero:2015, Gaggero:2017jts} in their KRA$\gamma$ CR propagation model.

Following our previously adopted convention, we refer with Case B to the calculations performed by assuming uniform CR spectral index, while we indicate with Case C the calculations performed by assuming that spectral index depends on the Galactocentric distance according to Eq.~\ref{Eq:Delta}.

\section{Neutrinos from sources}
\label{Neutrinos from sources}
In this paper, we take advantage of the constrain obtained in \cite{Cataldo:2020qla} on the total gamma-ray flux produced by all the sources (resolved and not resolved) in our Galaxy using the observations provided by the HGPS \citep{H.E.S.S.:2018zkf} in the $1-100$~TeV energy to derive an upper limit on the neutrino contribution from Galactic sources.
We assume that all gamma rays are produced by hadronic interaction and, hence, a neutrino counterpart is also expected.
Namely, we normalize the neutrino flux in order to reproduce the predicted total gamma-ray flux in the 1-100 TeV energy range.

\subsection{Total gamma-ray source flux}
In order to predict the cumulative gamma-rays and neutrino source signal, we follow the approach of \cite{Cataldo:2020qla}.
In particular, the source spatial and luminosity distribution is described as the product:
\begin{equation}
\frac{dN}{d^3 r\,dL} = \rho\left({\bf r} \right) Y \left(L\right)  
\label{SpaceLumDist}
\end{equation}
where ${\bf r}$ indicates the source position and $L$ is the source gamma-ray intrinsic luminosity integrated in the $1-100\, {\rm TeV}$ energy range probed by H.E.S.S.. 
The function $\rho({\bf r})$ is proportional to the pulsar distribution parameterized by \cite{Lorimer:2006qs} and scales as $\exp \left(-\left|z  \right|/H\right)$ with $H=0.2\ {\rm kpc}$, along the direction $z$ perpendicular to the Galactic plane.
It is conventionally normalized to one when integrated in the entire Galaxy. 
The function $Y(L)$ represents the source luminosity function and it is described by:
\begin{equation}
Y(L)=\frac{\mathcal N}{\LMax}\left(\frac{L}{\LMax}\right)^{-\alpha}
\label{LumDist1} 
\end{equation}
in the luminosity range $\LMin\le L \le
\LMax$.
In the above relation, $\LMax$ and $\mathcal N$ are the maximum TeV gamma-ray luminosity of the population and the high-luminosity normalization of the luminosity function, respectively.
It is also useful to introduce the maximum TeV emissivity, defined as:
\begin{equation}
\FMax = \frac{\LMax}{\langle E \rangle}
\end{equation}
where $\langle E \rangle$ is the average energy of photons emitted in the range $1-100\,{\rm TeV}$.
The total TeV gamma-ray flux produced by all the sources (resolved and not resolved) in the ANTARES observational windows is calculated by using the prescription of \cite{Cataldo:2020qla}:
\begin{equation}
\Phi_{\rm tot} = 
\xi\;
\frac{ \mathcal N \FMax}{4\pi (2-\alpha)}\; 
\langle r^{-2} \rangle
\label{phitot}
\end{equation}
where the parameter $\xi$, which is defined as
\begin{equation}
\xi \equiv \int _{\rm OW}d^3r \, \rho({\bf r}) = 0.48,
\end{equation}
represents the fraction of sources of the considered population which are included in the ANTARES observational window (OW), while the quantity $\langle r^{-2} \rangle$, defined as:
\begin{equation}
\langle r^{-2} \rangle \equiv \frac{1}{\xi}
\int_{\rm OW}d^3r \, \rho({\bf r}) \; r^{-2} = 0.013 \,{\rm kpc}^{-2}
\end{equation}
is the average value of their inverse square distance. 
In order to calculate the total flux, we use the best-fit values for the maximum luminosity $\LMax$ and the high-energy normalization $\mathcal{N}$ derived in \cite{Cataldo:2020qla} by fitting the flux, latitude, and longitude distribution of bright sources in the HGPS catalog.
In particular, we take the values $\LMaxTeV =  5.1^{+3.4}_{-2.2} \times 10^{35}{\rm erg\;s^{-1}}$ and $\mathcal{N} = 18^{+14}_{-7}$ that are obtained for $\alpha=1.5$ by considering the entire sample of 32 sources above the H.E.S.S. completeness threshold\footnote{The compleatness threshold is $\Phi_{\rm th}=0.1 \Phi_{\rm CRAB}$ where $\Phi_{\rm CRAB}=2.26\times 10^{-11} {\rm cm^{-2}\,s^{-1}}$ is the flux of the CRAB nebula integrated above $1$ TeV \citep{H.E.S.S.:2018zkf}.} in the HGPS catalog. 
The above results are obtained by assuming that all the gamma-ray sources have a power-law spectrum with a spectral index equal to $2.3$ \citep{H.E.S.S.:2018zkf}.
This corresponds to assuming $\langle E \rangle = 3.25$ TeV from which it follows that the best-fit maximal emissivity of the population is $\FMax = 9.9^{+6.5}_{-4.2} \times 10^{34}\,{\rm s^{-1}}$.
If we consider a different spectral assumption, the best-fit value of $\mathcal{N}$ remains unchanged while $\LMax$ is shifted proportionally to the variation of $\langle E \rangle$. As a consequence, the best-fit value of $\FMax$ remains constant and the total integrated gamma-ray flux $\Phi_{\rm tot}$ is unchanged, as can be understood by considering Eq.~\ref{phitot}.  


%
%
%

\subsection{Total neutrino source flux.}

The total neutrino flux is obtained from the gamma-ray flux in the following way. We assume that the CR injected spectrum can be parameterized as: 
\begin{equation}
\phi_{p}(E) = \frac{1}{K_{\rm p}} \left(\frac{E}{1\, {\rm  TeV}} \right)^{-\beta}\exp{\left(-\frac{E}{E_{\rm cut}}\right)}.
\label{CRspectrum}
\end{equation}
where $K_{\rm p}$ is a suitably defined normalization constant (whose value is not relevant to our calculations). 
Hence, the gamma-ray and the all-flavor neutrino spectra (normalized to 1 in the $1-100\,{\rm TeV}$ energy window) produced by hadronic interaction within the source can be calculated in the following way:
\begin{eqnarray} \label{gamma and neutrino source flux}
\phi_{\gamma}(E_{\gamma}) &=&
 \frac{1}{K_\gamma}\int_{E_{\gamma}}^{\infty}
dE\, \frac{d\sigma(E,E_{\gamma})}{dE_{\gamma}}\,
\phi_{p}(E) \\
\nonumber
\phi_{\nu}(E_{\nu}) &=&  \frac{1}{K_\nu}\left[\sum_{l=e,\mu,\tau}
\int_{E_{\nu}}^{\infty} dE\, \frac{d
\sigma_{l}(E,E_{\nu})}{dE_{\nu}}
\phi_{p}(E) \right]
\end{eqnarray}
where $\frac{d \sigma(E, E_{\gamma})}{dE_{\gamma}}$ ($\frac{d\sigma_l(E, E_{\nu})}{dE_{\nu}}$) represents the differential cross section for the production of gamma-ray (neutrino with flavor $l$) by a nucleon of energy $E$ in nucleon-nucleon collisions, as parameterized in \cite{Kelner:2006tc}. 
The normalization constants $K_\gamma$ and $K_\nu$ are given by:
\begin{eqnarray} \label{gamma and neutrino nromalizations}
K_\gamma &=& \int^{E_ {\rm sup}}_{E_{\rm inf}} dE_\gamma
\int_{E_{\gamma}}^{\infty}  dE\,
\frac{d\sigma(E,E_{\gamma})}{dE_{\gamma}}\,
\phi_{p}(E) \\
\nonumber
K_\nu &=&  \sum_{l=e,\mu,\tau} \int^{E_ {\rm sup}}_{E_{\rm inf}} dE_\nu
\int_{E_{\nu}}^{\infty}  dE\, \frac{d
\sigma_{l}(E,E_{\nu})}{dE_{\nu}} \, \phi_{p}(E)
\end{eqnarray}
where $E_{\rm inf} = 1\,{\rm TeV}$ and $E_{\rm sup} = 100\,{\rm TeV}$ define the boundaries of the energy region probed in gamma-rays by the H.E.S.S. detector. The quantity:
\begin{equation}
\eta \equiv \frac{K_{\nu}}{K_{\gamma}}
\end{equation}
represents the ratio between the number of neutrinos (of all flavor) and the number of photons that a given source produce in the energy window $[E_{\rm inf}, E_{\rm sup}]$. This ratio is only determined by the assumed injected CR spectral shape, while it is independent of the column depth of target material within the source, the source distance and intrinsic luminosity, etc.

By using the above definitions, we are able to take advantage of the bounds on the total gamma-ray flux derived by \cite{Cataldo:2020qla} to calculate the cumulative neutrino emission produced by all sources (resolved or not) contained in a given observation window. The all-flavor differential neutrino flux produced by sources is indeed given by:
\begin{equation}
 \varphi_{\nu, s} (E_\nu) = \Phi_{\nu, \rm tot} \,  \phi_\nu (E_\nu)
\end{equation}
where $\Phi_{\nu, {\rm tot}} \equiv \eta \, \Phi_{\rm tot}$ represents the cumulative neutrino flux integrated in the energy window $[E_{\rm inf}, E_{\rm sup}]$.




The neutrino flux is calculated under different assumptions for the CR injected spectrum.
In particular, we fix $\beta=2.4$ in order to reproduce the average measured gamma-ray spectral index that is $2.3$ \cite{H.E.S.S.:2018zkf}. Steeper spectra than $E^{-2}$, which is the standard result for the CR injected spectrum in case of diffusive shock acceleration~\footnote{The general result is given in momentum $p^{-4}$ that becomes $E^{-2}$ in case of relativistic particles.}, are also justified by theoretical arguments, see, e.g., \cite{Caprioli:2020spz}.
Concerning the cutoff energy, we consider two different values $E_{\rm cut}=0.5-10$~PeV to explore the relevance of this parameter for our final results.
%

\section{Results}
\label{Results}

\begin{figure}[h!]
\begin{center}
\includegraphics[width=0.8\textwidth]{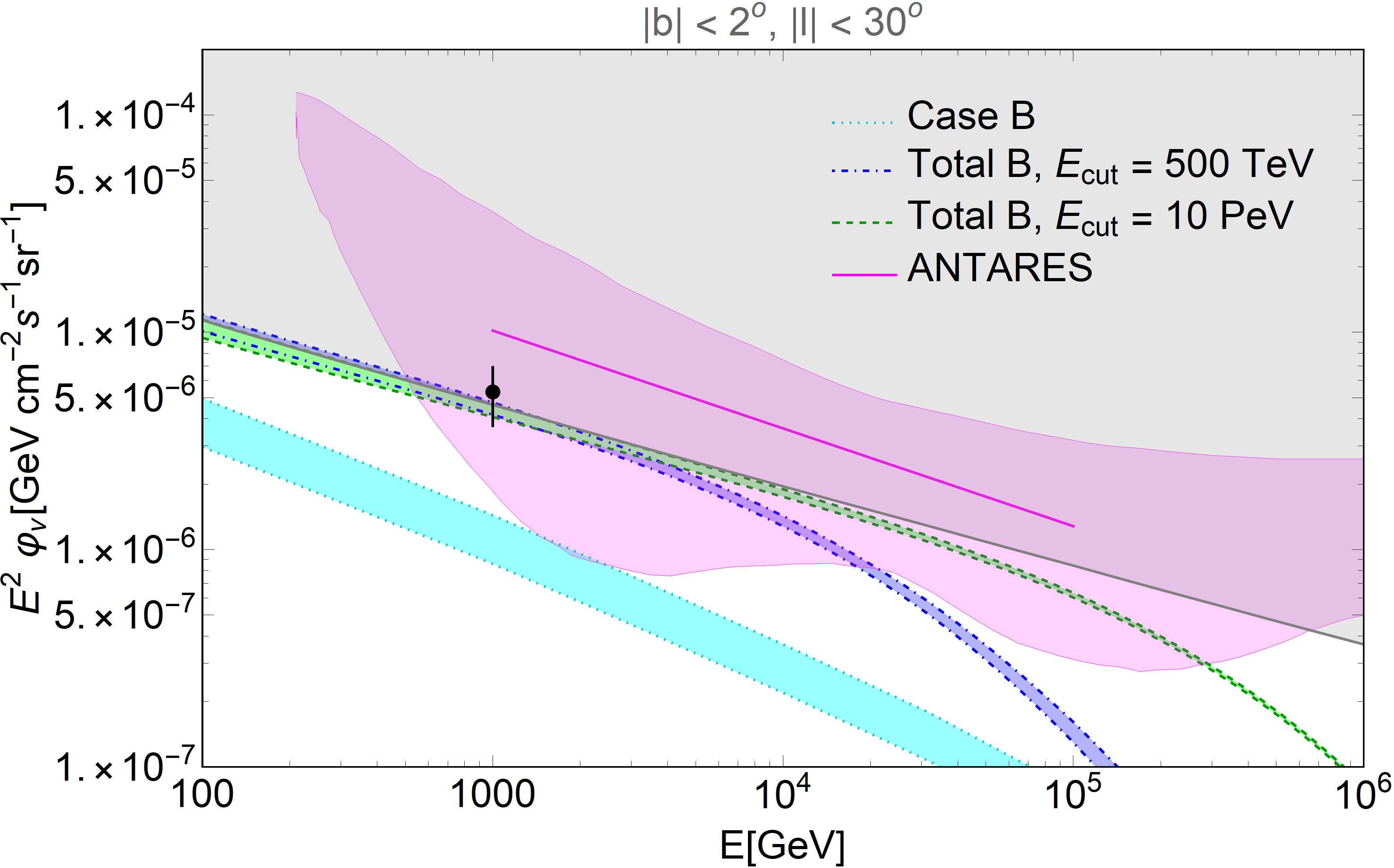}
\caption{\small\em Differential energy spectra of diffuse neutrino from the Galactic plane in the angular regions probed by ANTARES. 
The magenta line corresponds to the ANTARES best fit for neutrino from the Galactic ridge. The magenta band corresponds to the $1\sigma$ uncertainties.
The predictions for the diffuse emission (Case B) are shown with a cyan (dotted) band.
The predictions for the total neutrino flux (Case B + sources) are shown with blue (dot-dashed) and green (dashed) bands, corresponding to energy cut $E_{\rm cut}=500$ TeV and $E_{\rm cut}=10$ PeV, respectively. 
The bands represent the uncertainties on the spatial distributions of CRs in our Galaxy. The lower (upper) limit of the band is obtained by assuming a smearing radius equal to 1 kpc (infinity).
We show the effect of different energy cutoffs for the CR source spectra as displayed in the labels.
We additionally display an excluded region in gray. The bottom line corresponds to the maximum neutrino contribution from our Galaxy obtained by assuming Case B for the diffuse emission and a power law for the CR source spectra with index 2.4.
 The black point represents the total neutrino flux obtained by converting the total gamma-ray signal measured by the H.E.S.S. experiment at 1 TeV. 
 The error bar includes the $30\%$ systematic uncertainty on the flux \citep{H.E.S.S.:2018zkf}.}
\label{fig:AntaresB}
\end{center}
\end{figure} 

\begin{figure}
\begin{center}
\includegraphics[width=0.8\textwidth]{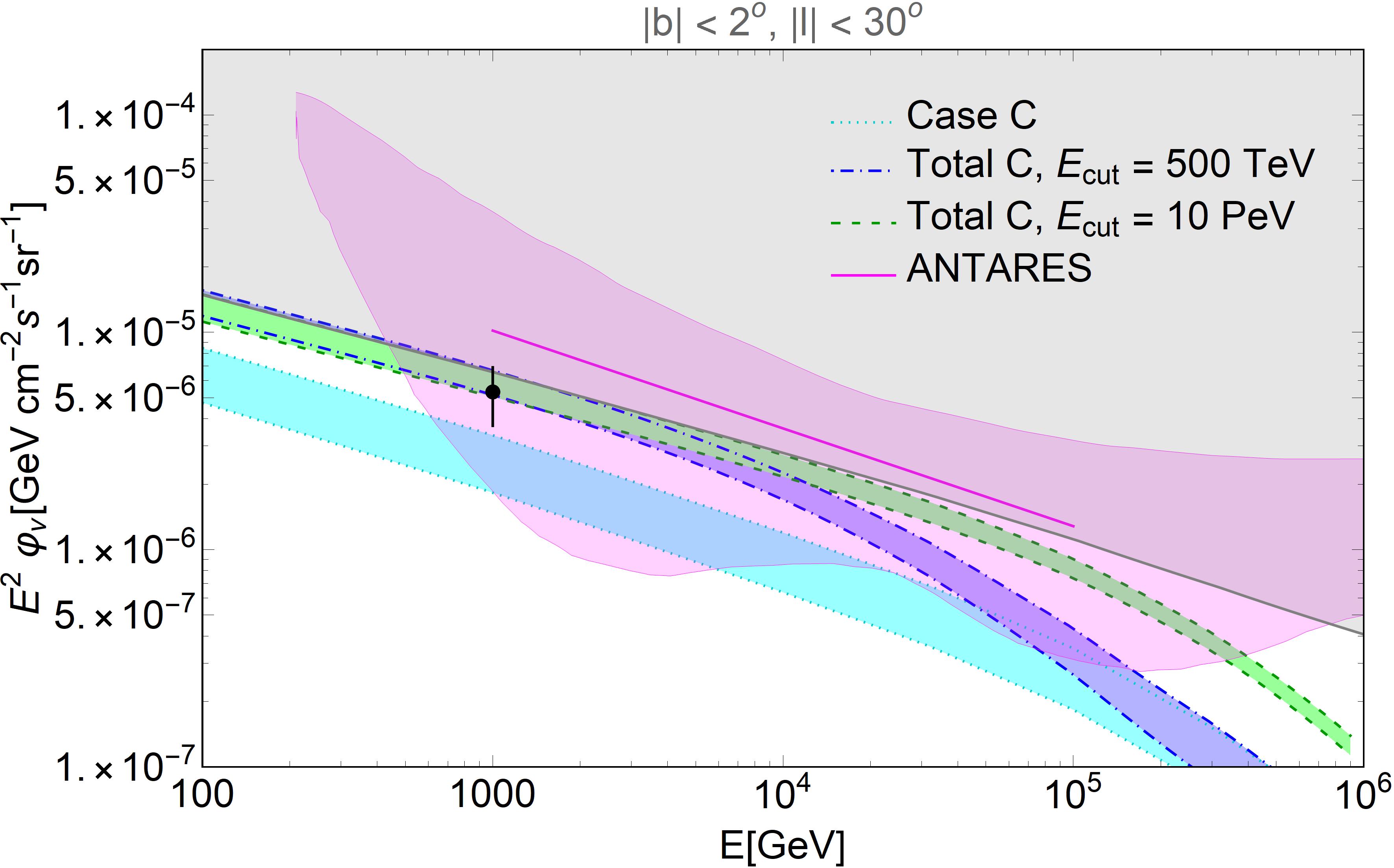}
\caption{\small\em Same as Fig.~\ref{fig:AntaresB} but calculated considering Case C for the diffuse neutrino emission.}
\label{fig:AntaresC}
\end{center}
\end{figure}

\subsection{Comparison with ANTARES}

Our predictions for the all-flavor Galactic neutrino flux in the ANTARES observational window are shown in Fig.~\ref{fig:AntaresB} and \ref{fig:AntaresC}.
The cyan bands in the figures correspond to diffuse neutrino emission
calculated assuming uniform (Case B) and position-dependent (Case C)
CR spectral index, respectively. 
The total flux, obtained as the sum of the neutrino diffuse and source emission, is displayed with blue and green bands for two different values
$E_{\rm cut}= 0.5,\,10$~PeV of the energy cutoff of the primary proton source spectra, as it is indicated in the labels.
The width of the displayed bands is related to the
uncertainties in the assumed CR spatial distribution in our Galaxy.
In particular, the lower limit of each band is obtained assuming the
diffusion length $R=\infty$, while the upper limit is obtained assuming $R=1$ kpc.
Finally, the ANTARES best-fit flux, obtained in the $1-100$ TeV
energy range, is shown with a magenta line. The magenta band represents the $1\sigma$ uncertainties \citep{ANTARES:2022izu}.

Let us discuss first the pure diffuse contribution due to the interaction of CR with the interstellar medium. 
We can see from Fig.~\ref{fig:AntaresB} and~\ref{fig:AntaresC} that the predicted diffuse neutrino emission is well below the signal potentially observed by ANTARES.
When we consider the standard assumption, i.e., our Case B, the diffuse neutrino flux at the energy  $E_\nu \sim 40\,{\rm   TeV}$ which is most efficiently probed by ANTARES, is a factor $\sim 10-20$ lower than the potentially observed signal.
%
We remark that, in this scenario, the diffuse component is relatively
well constrained.
Since the CR spectral distribution is assumed to be position-independent, the expected diffuse neutrino and gamma-ray emission spectra are indeed universal.
In other words, the neutrino and gamma-ray fluxes from a given sky direction have fixed spectral shapes and only depend on normalization (energy-independent) factors that are proportional to the product $n_{\rm H}({\bf r}) \, g({\bf r})$ integrated along the line of sight.
These normalizing factors can be constrained by gamma-ray observation performed in different energy bands.
Thus,  we cannot expect larger uncertainties than those already introduced by letting the diffusion radius $R$ vary, as it is described in Sect.~\ref{Neutrino diffuse flux}.

The only possibility to increase the expected diffuse neutrino flux is to assume that the CR spectrum in the inner Galaxy is harder than at the Sun position,  as implemented in our Case C by introducing the function $h(E,{\bf r})$ suitably defined to reproduce the results in \cite{Gaggero:2015,Yang:2016jda, Pothast:2018bvh} based on Fermi-LAT large-scale $\gamma$-ray data in the GeV energy domain.
Even in this upper scenario, however, the predicted diffuse neutrino emission is well below ANTARES signal, being a factor $\sim 3-6$ lower than the best-fit flux at $E_\nu \sim 40\,{\rm  TeV}$.
It is also worth remarking in the context of this scenario that the total TeV gamma-ray data could be in slight tension with this assumption that saturates the observed emission \citep{Cataldo:2019qnz}. Moreover,    
a fraction of the spectral index variation of the large-scale gamma-ray emission observed by Fermi-LAT may be naturally explained by a population of unresolved PWNe, see \cite{Vecchiotti:2021vxp}, thus weakening the evidence in favor of CR spectral hardening in the inner Galaxy.


The above discussion allows us to conclude that the ANTARES hint for a Galactic neutrino signal cannot be explained by CR diffuse emission alone and requires a dominant contribution from a population of Galactic neutrino sources.
The source contribution, moreover, should be very large, as it can be immediately understood by comparing our predictions for the total (diffuse + source) neutrino flux with the ANTARES signal.
We see indeed that the blue and green bands in the Figures are always below the ANTARES best-fit results.
We recall that the source population considered in this study is constrained to reproduce the flux, longitude, and latitude distribution of $\gamma-$ray emitting objects observed by HGPS in the TeV energy domain. 
Moreover, all the sources are assumed to emit radiation by hadronic interactions.
This second assumption is rather extreme and allows us to obtain a very conservative upper bound for the source neutrino contribution.
Indeed, according to our present knowledge, the TeV gamma-ray Sky seems to be dominated by Pulsar Wind Nebulae that are mostly powered at TeV by the IC mechanism, with subdominant or negligible neutrino emission.
We see from the figures that, in order to have a total neutrino emission comparable to the ANTARES best-fit flux, in addition to the above hypotheses, it is also necessary to postulate that all these sources are Pevatrons, i.e., they accelerate protons at energy larger than $\sim 1$~PeV.

Our calculations are finally used to set an upper limit for the total neutrino flux from the disk of our Galaxy. 
The solid grey lines show the maximal predictions for the total neutrino flux that can be obtained in the considered scenarios.
They are indeed obtained by maximizing both diffuse and source components, i.e., by taking $R=1\,{\rm kpc}$ for diffuse flux calculation and $E_{\rm cut}= \infty$ for the primary proton source spectrum.
We also show with a black point the total neutrino flux obtained by converting the total gamma-ray flux measured by the H.E.S.S. experiment at 1 TeV integrated into the ANTARES observational windows using the spectral assumption for the CR given in Eq.~\ref{CRspectrum}. 
The H.E.S.S. experiment provides a measurement of the spatial profile of the total gamma-ray emission, averaged over latitudes $|b| < 2^\circ$, in the longitude range  $-75^\circ < l < 60^\circ$, for a photon median energy $E_\gamma = 1\,$TeV \citep{Abramowski:2014}\footnote{The total emission from the Galactic plane is obtained as the excess with respect to the average signal at absolute latitudes $|b|\ge 1.2^\circ$. This background subtraction procedure cancels out all the signals with a large latitudinal profile, e.g., the IC scattering contribution.}.
The total flux measured by H.E.S.S. includes the contribution from sources and diffuse emissions, and it has to be interpreted as an upper limit on the neutrino flux expected at 1 TeV. The error bar represents the systematic error on the flux that is of order $30\%$ \citep{H.E.S.S.:2018zkf}.
We see that the best-fit flux and a relevant part of the ANTARES $1\sigma$ region lie above the maximal allowed neutrino emission according to our calculations and/or H.E.S.S. $\gamma-$ray data.

The observation of a Galactic neutrino signal inside the gray shaded regions in Fig.~\ref{fig:AntaresB} and \ref{fig:AntaresC} basically corresponds to assuming that the Galactic disk is much brighter in neutrinos than in photons.
Since neutrino emission is always accompanied by a comparable production of photons in the hadronic mechanism, this would require that the sources responsible for the neutrino production are opaque in gamma-ray. 
In this scenario, since the gamma-ray signal is absorbed inside sources, it would not be possible to correlate the gamma-ray and neutrino observations in a straightforward way.

\section{Conclusions}
\label{Conclusions}
Recently, the ANTARES collaboration reported the first hints towards the existence of neutrino flux from the Galactic ridge in the angular region $|l|<30^{\circ}$ and $|b|< 2^{\circ}$ and in the $1-100$ TeV energy band \citep{ANTARES:2022izu}.
In this work, we predict the total neutrino flux expected from the disk of our Galaxy. 
In particular, we include the contribution of the diffuse emission, produced by the interaction of CRs with the ISM, and Galactic sources.
The diffuse emission is calculated under different assumptions for the CR spatial and energy distribution. We considered two cases: Case B which assumes the CRs spectrum is the same everywhere in the Galactic disk, and Case C which assumes that the CR spectral index depends on the Galactocentric distance.
We showed that also in the most optimistic Case C, the solely diffuse emission is not able to explain the ANTARES observation being a factor $\sim3-6$ lower than the ANTARES best fit.
A source component is required to explain the neutrino observation.
Hence, we estimated the neutrino contribution from Galactic sources using gamma-ray observations.
In particular, we considered the total gamma-ray flux, as constrained by \cite{Cataldo:2020qla} from analysis of flux, longitude, and latitude distributions of the brightest sources (above the completeness threshold) of the HGPS catalog \cite{H.E.S.S.:2018zkf}.
We derived a very conservative upper bound for the neutrino source contribution by assuming that all the gamma rays are produced by hadronic mechanisms.
We compared the total neutrino flux, given by the sum of diffuse and source emission, with the ANTARES best fit.
Even in this case, our total signal is below the best-fit value.
As a last step, we set an upper limit for the total neutrino flux by maximizing both the diffuse emission, by taking $R=1\,{\rm kpc}$, and the source components, by assuming $E_{\rm cut}= \infty$ for the primary proton source spectrum.
In the near future, new measurements from current and upcoming neutrino detectors will be able to constrain the fraction of Galactic sources that contribute to the neutrino signal from the Galactic disk. If new observations lie above the derived upper limit, our result would imply that neutrinos are produced by sources that are opaque in gamma rays.


\acknowledgments
The work of VV is supported by the European Research Council (ERC) under the ERC-2020-COG ERC Consolidator Grant (Grant agreement No.101002352).
The work of GP and FLV is partially supported by the research grant number 2017W4HA7S ''NAT-NET:
Neutrino and Astroparticle Theory Network'' under the program PRIN 2017 funded by the Italian Ministero dell'Istruzione, dell'Universita' e della Ricerca (MIUR).

\paragraph{Note added.} After the conclusion of this work and its submission to arXiv, IceCube reported the observation of high-energy neutrinos from the Galactic plane \citep{IceCubeScience} at 4.5$\sigma$ level significance. This makes even more timely the evaluation of the expected Galactic Ridge neutrino signal performed in our paper. Unfortunately, the IceCube results are based on template fitting which relies on specific assumptions for the assumed angular and energy distribution of the expected signal. As a consequence, the comparison with our calculations and with the hint reported by ANTARES is not straightforward and it will be considered in a future publication.


\bibliographystyle{JHEP}
\bibliography{bibliography.bib}


\end{document}